\documentclass[12pt]{article}
\usepackage[pdftex]{graphicx}
\usepackage{caption,tabularx,booktabs}
\usepackage[pdf]{pstricks}
\usepackage{amsmath}
\usepackage{pgffor}
\usepackage{pdflscape,tikz}
\usepackage{longtable,subfig,placeins}
\usepackage[onehalfspacing]{setspace}
\usepackage{amsmath, amsthm,natbib}
\usepackage{amssymb}
\usepackage{epstopdf}
\usepackage{csquotes}
\usepackage{epsfig}
\usepackage{amsfonts}
\usepackage{rotating}
\usepackage[colorlinks, citecolor=blue, urlcolor=blue, pdfborder={0 0 1}, citebordercolor={1 1 1}]{hyperref}
\usepackage{hyperref}
\usepackage{amsfonts}
\usepackage{booktabs}
\usepackage{siunitx}
\usepackage{float}
\usepackage{footmisc}
\usepackage[ruled]{algorithm2e}
\usepackage{enumitem}
\usepackage{dcolumn}
\usepackage{fancyvrb}
\usepackage{lipsum}
\usepackage{lmodern}
\usepackage[T1]{fontenc}

\usepackage{footnote}

\usepackage{ragged2e}
\newcolumntype{P}[1]{>{\RaggedRight\arraybackslash}p{#1}}

\usetikzlibrary{bayesnet}
\newsavebox{\tempbox}

\begin{document}
\onehalfspacing

\title{\textbf{Surveying Generative AI's Economic Expectations}
\thanks{leland.bybee@yale.edu. 
I benefited from discussions with and comments from Nick Barberis, Paul Fontanier, Will Goetzmann, Paul Goldsmith-Pinkham, Bryan Kelly, Alp Simsek, Kaushik Vasudevan
as well as seminar participants at the Yale SOM brown bag.
}
}

\author{J. Leland Bybee\\ \footnotesize Yale University\\}
\date{\footnotesize First Draft: February 16, 2023 \\
This Draft: April 26, 2023 \\
\emph{Preliminary, comments welcome}}

\maketitle
\begin{abstract}
I introduce a survey of economic expectations formed by querying a large language model (LLM)'s expectations of various financial and macroeconomic variables based on a sample of news articles from the Wall Street Journal between 1984 and 2021.
I find the resulting expectations closely match existing surveys including the Survey of Professional Forecasters (SPF), the American Association of Individual Investors, and the Duke CFO Survey.
Importantly, I document that LLM based expectations match many of the deviations from full-information rational expectations exhibited in these existing survey series.
The LLM's macroeconomic expectations exhibit under-reaction commonly found in consensus SPF forecasts.
Additionally, its return expectations are extrapolative, disconnected from objective measures of expected returns, and negatively correlated with future realized returns.
Finally, using a sample of articles outside of the LLM's training period I find that the correlation with existing survey measures persists
-- indicating these results do not reflect memorization but generalization on the part of the LLM.
My results provide evidence for the potential of LLMs to help us better understand human beliefs and navigate possible models of nonrational expectations.

%

\end{abstract}

\thispagestyle{empty}

\newpage

\setcounter{page}{1}
\begin{displayquote}
    \textit{
Expectations matter. How much to consume or save, what price to set, and whether
to hire or fire workers are just some of the fundamental decisions underlying macroeconomic dynamics that hinge upon agents’ expectations of the future. Yet how
    those expectations are formed, and how best to model this process, remains an open question.}
\begin{flushright}
    \cite{Coibion2015Aug}
\end{flushright}
\end{displayquote}

\begin{displayquote}
    \textit{
Beliefs are central to asset pricing.
Asset prices are forward-looking, and essentially any asset-pricing model implies that investors price assets based
on their beliefs about the joint distribution of some stochastic discount factor (SDF) $M_{t+1}$ and payoffs $X_{t+1}$.
An observer outside the field of asset pricing might therefore guess that a major part of the research efforts in asset pricing
are devoted to understanding how investors form beliefs.
    This is, at least so far, not the case.}
\begin{flushright}
    \cite{Brunnermeier2021Apr}
\end{flushright}
\end{displayquote}

\begin{displayquote}
    \textit{
        Anyone who sets himself the problem of analyzing the causation of behavior will (in the absence of independent neurophysiological evidence) concern himself with the only data available, namely the record of inputs to the organism and the organism's present response.}
    \begin{flushright}
        \cite{chomsky1959review}\footnote{This quote is taken from Chomsky's review of \cite{skinner1957verbal} and presents Chomsky's summary of Skinner's approach.  While the review itself is critical, the behaviorist approach, so summarized, is nevertheless relevant to the present study of LLMs.}
    \end{flushright}
\end{displayquote}

\newpage

\section{Introduction}

The rational expectations hypothesis of \cite{muth1961rational} remains the dominant model of beliefs in macroeconomics and finance.
Its dominance is not hard to understand: rational expectations provides a tractable approach for modeling economic behavior (\cite{Lucas1972Apr}),
while constraining the econometrician's degrees of freedom (\cite{lucas1976econometric}).
However, rational expectations has never been without evidence and alternative theories to question its dominance.
Such models place us in the ``wilderness'' of alternative expectations of \cite{sims1980macroeconomics}.\footnote{Sims himself refers to the ``wilderness of disequilibrium economics'' but this wilderness has since been rephrased as \cite{sargent2001conquest}'s ``wilderness of bounded rationality'' and \cite{Angeletos2021Jan}'s ``wilderness of alternative models for expectations formation and equilibrium''.}
Successful attempts to navigate this wilderness have primarily relied on survey data for proxies of beliefs (\cite{Coibion2015Aug}, \cite{bordalo2020overreaction}, \cite{Angeletos2021Jan}, \cite{Nagel2022May}, \cite{Lochstoer2022Apr}).

In this paper I propose a new source for proxies of beliefs using large language models (LLMs).
LLMs, are a class of statistical model designed to learn the structure of human language.
These models accomplish this goal by estimating the probability of a token, $s_i$, given all previously observed tokens in a document,

\begin{equation}
    p(s_i|s_1, \ldots, s_{i-1}),
\end{equation}
using a particular neural network architecture, known as transformers (\cite{Vaswani2017Jun}).
While this objective may seem simple, by leveraging massive corpuses of training data and a very large parameter space, these models have exhibited an emergent ability to mimic ``human-like'' behavior (\cite{Brown2020May}, \cite{Wei2022Jun}, \cite{Bubeck2023Mar}).\footnote{Emergent here refers to the scale of these models as opposed to smaller language models.  For instance, Open AI's GPT-3 instance used roughly 45 terabytes of text for training data and has roughly 175 billion parameters.}

This ability to mimic human-like interactions with text opens up the possibility for LLMs to simulate human behavior and beliefs.
Rich causal beliefs can be embedded in text -- this assumption underlies the entire academic publication process.
As a statistical procedure designed to generate text similar to its training corpus, LLMs should reflect the beliefs embedded in their training data.\footnote{That LLMs reflect the biases and beliefs of their training corpus is central to much of the current work on AI alignment (\cite{Bender2021Mar}, \cite{Schramowski2022Mar})}
I leverage this insight to form expectations of an LLM by providing a historical sample of news articles from \emph{The Wall Street Journal} (\emph{WSJ}) to OpenAI's GPT-3.5 instance and asking it to predict various financial and macroeconomic quantities.
I then aggregate these article level expectations into a time-series of monthly and quarterly expectations and compare the resulting belief proxies to existing surveys.

First, I validate how well GPT's expectations of the stock market match the expectations of the American Association of Individual Investors (AAII) survey and Duke CFO Survey.
I  find GPT's expectations significantly correlate with these existing surveys on a comparable level to alternative existing survey measures.
I then compare the correlation between GPT's expectations of the S\&P and various alternative measures of expected returns to that of the existing surveys.
I find that GPT exhibits the same extrapolative expectations and significantly correlate with equity fund flows.
Additionally, I find that GPT's expectations exhibit the same disconnect from objective measures of expected returns -- matching the sign of the correlation for existing survey measures with the log dividend price ratio, \cite{Lettau2001Jun}'s CAY measure, and several predictive proxies for expected returns.
Finally, I find that GPT's expectations are negatively correlated with future realized returns, matching existing survey measures while differing from objective measures of expected returns.

Second, I compare GPT's expectations, for a series of macroeconomic variables studied in \cite{Coibion2015Aug}, to the Survey of Professional Forecaster (SPF).
I find that GPT's expectations are significantly correlated with revisions in all but two of the SPF series.
I then evaluate whether the variation in SPF revisions associated with GPT's expectations is a driver of the underreaction commonly found in consensus SPF forecasts.
By running Coibion-Gorodnichenko (CG) regressions using both SPF revisions and GPT's expectations, I find that GPT is able to match the observed underreaction.

These results provides new evidence for navigating the wilderness of alternative expectations as GPT's expectations match many of the deviations from full-information rational expectations (FIRE) exhibited in existing survey series.
There are two possible sources for deviations from FIRE in GPT's expectations, $F_t^{gpt}$, about the $k$th series, $X_{t+h}^k$, given as follows:

\begin{equation}
    F^{gpt}_{t}(X^k_{t+h}) = g_k(\theta_t).
\end{equation}
Here $g_k$ corresponds to the estimated LLM weights prompted for the given series $k$ and $\theta_t$ the provided news text at period $t$.
First, note that $g_k$ is static across the entire sample.
Second, note that the news text provided to GPT is isolated and lacking in context -- GPT's expectations at time $t$ are based only on the set of news articles available at time $t$.\footnote{In fact it is stronger than this, GPT's expectations are based on individual news articles and even lack the context of other news articles at time $t$.}
This puts constraints on the set of models which can explain GPT's deviations.

If the deviations in GPT's expectations emerge from the estimated LLM weights, $g_k$, then models to explain then models of beliefs must reflect a static reaction function to news.
Models that rely on an overemphasis of representative or salient news (\cite{bordalo2018diagnostic}, \cite{Bordalo2022Aug}) may explain this bias if the representativeness is not dependent on recent context but some more general attribute of the given news articles.
The potential of the LLMs to capture these biases indicates the importance of a nascent multidisciplinary literature experimentally studying the behavior of these models (\cite{aher2022using}, \cite{Horton2023Apr}, \cite{Argyle2023}, \cite{brand2023using}).


Alternatively, if deviations in GPT's expectations emerge from attributes of the news articles themselves, $\theta_t$, then models to explain these deviations need to address these attributes.
In particular, for models of learning, extrapolation, and the representativeness of recent states to explain these deviations (\cite{Fuster2010Dec}, \cite{afrouzi2020overreaction}, \cite{farmer2021learning}), it must be the case that these dynamics are embedded in the news text generating process itself.
This potential source of deviations highlights the importance of work studying the origin and dynamics of narratives as a driver of beliefs (\cite{Shiller2019Jan}, \cite{bybee2021business}, \cite{andre2021}, \cite{Michalopoulos2021Nov}, \cite{bybee2022narrative}, \cite{flynn2022}).

Further still, it is interesting to note that GPT is able to match distinct moments across a variety of existing survey series.
Given GPT's expectations are based on the same algorithm and the same source of news, this suggest the distinct deviations observed across these surveys originate from the same source and models attempting to explain these moments should be able to jointly explain these results.

An alternative story that may contaminate my results is the possibility of data leakage from the training corpus.
Importantly for my story, I assume that GPT has learned some generalization of the causal structure of beliefs contained in its training sample not that it is responding with memorized text.
To address the relevance of this concern for my results I run two tests.
First, using a sample of \emph{WSJ} articles outside of GPT's training period -- after September of 2021 -- I test whether GPT's expectations continue to correlate with the existing survey measures, I find that this is indeed the case.
Second, if GPT is relying on information about the future then realized values of the corresponding series should be more important for explaining GPT's expectations than the forecast revisions.
I find that this is not the case.

An open question still remains: who's beliefs are GPT's?
My results suggest GPT's beliefs closely match a number of different groups, including professional forecasters, individual investors, and CFOs.
GPT's training corpus is based on a wide range of sources and topics, and as such it may provide something of a ``representative agent''.
Nevertheless, LLMs designed to more directly reflect the beliefs and interest of distinct groups is likely the future of this field.
This work may be done by fine tuning existing open source LLMs like that in \cite{Touvron2023Feb} or building from the ground up from distinct corpuses as in \cite{Wu2023Mar}.
My results provide a proof of concept that these methods represent a valuable route forward to better understand human behavior.


%

\section{Measuring Generative AI's Expectations}

My primary dataset consists of all articles contained in \emph{The Wall Street Journal} from 1984 to 2021 purchased from the Dow Jones Historical News Archive.
A full summary of the data cleaning procedure is detailed in appendix \ref{app:filtersmain}.
Beyond this initial cleaning I sample 300 articles randomly from each month of data.
This is done because each request made to OpenAI's API costs a marginal fee.


Queries are made to OpenAI's GPT-3.5 model instance.
Figure \ref{fig:prompt} reports the prompt format used to query GPT.
Each query to GPT receives a headline and a description of the desired series -- for instance, for CPI I will ask about an increase or decrease in ``the consumer price index in the United States''.
In response GPT generates a string of text corresponding to the requested format in the prompt.

\begin{figure}[!t]
    \caption{Prompt Format}
    \label{fig:prompt}
\begin{center}
    \input{core_question_format.txt}
\end{center}
    {\footnotesize {\it Note.}
    Reports the prompt format for queries made to GPT.
    ``\%s'' indicates where in the prompt the headline and target text are inserted.
    }
\end{figure}

Table \ref{tab:sstats} reports the surveyed series as well as summary statistics of the responses.
Figure \ref{fig:coocurr} in appendix \ref{app:coocurr} reports the coocurrence between the possible pairings of increase/decrease across the surveyed series.\footnote{As opposed to classical supervised machine learning, GPT is not trained on a specific task, i.e. forecasting economic quantities, but rather on a general language modeling task.
As such, this sort of classification is referred to as zero-shot learning in the machine learning literature.}
Table \ref{tab:example} reports a series of example headlines and the corresponding responses for the S\&P 500, CPI, and unemployment.
Additionally, I request GPT to provide an explanation for its response which is included in table \ref{tab:example}.
These explanations provide a useful method to analyze the reasoning behind GPT's responses.
For instance, when presented with a news article about the opening of access to the Japanese phone market for U.S. firms, GPT expects this will increase the S\&P by providing growth opportunities for U.S. phone companies, will decrease unemployment by creating new job opportunities in the U.S. telecommunications industry, and lower the CPI by increasing competition.
Similarly, when presented with an article about tax cuts and tightening government budgets during the Bush years, GPT expects this will increase the S\&P by stimulating the U.S. economy, no impact unemployment due to uncertainty about which effect may dominate, and increase the CPI by increased demand as a result of the tax cuts.
These explanations highlight GPT's ability to provide a clear chain of reasoning for its answers and open up new possible routes for understanding the causal structure of beliefs.

\begin{table}[!t]
    \begin{center}
    \caption{Summary Statistics of GPT Survey}
    \label{tab:sstats}
        \begin{tabular}{lp{0.35\textwidth}ccccc}
    \toprule
            Series & Prompt & Date Range & Count & Inc. \% & Dec. \% & Unc. \% \\
    \midrule
         SNP &                                           the S\&P 500 index &  1984-2021 & 136345 & 15.13 & 26.84 & 58.02 \\
 CPI &               the consumer price index in the United States &  1984-2021 & 132736 &  7.86 &  6.45 & 85.69 \\
  HS &                         housing starts in the United States &  1984-2021 & 132212 &  2.50 &  5.68 & 91.82 \\
  IP &                  industrial production in the United States &  1984-2021 & 132892 & 10.11 & 11.72 & 78.17 \\
DEFL &                 the GDP price deflator in the United States &  1984-2021 & 132760 &  9.63 & 12.50 & 77.88 \\
 AAA &              the AAA-rated bond's rate in the United States &  1984-2021 & 133467 & 11.09 & 14.88 & 74.03 \\
   C &                       real consumption in the United States &  1984-2021 & 131574 & 11.53 & 17.67 & 70.80 \\
  GF &         federal government consumption in the United States &  1984-2021 & 132839 &  9.86 & 10.88 & 79.26 \\
  GY &                           the real GDP of the United States &  1984-2021 & 132148 & 20.54 & 20.91 & 58.56 \\
 NRI &         real nonresidential investment in the United States &  1984-2021 & 132961 & 17.46 & 22.94 & 59.59 \\
  RI &            real residential investment in the United States &  1984-2021 & 133157 &  8.66 & 16.49 & 74.85 \\
  GS & state and local government consumption in the United States &  1984-2021 & 131428 & 13.44 & 17.30 & 69.26 \\
 3TB &                              the 3-month treasury bill rate &  1984-2021 & 134609 & 15.45 & 11.21 & 73.34 \\
  UE &                             employment in the United States &  1984-2021 & 120102 &  9.81 & 11.24 & 78.95 \\
        \\[-9pt]
        \bottomrule
    \end{tabular}
    \end{center}
    {\footnotesize {\it Note.}
    Reports summaries of the surveyed labels, the text passed to GPT, the date range, the number of non-missing responses,
    the proportion of responses that were ``increase'',
    the proportion of responses that were ``decrease'',
    and the proportions that were ``uncertain''.
    }
\end{table}

\begin{table}[!t]
    \begin{center}
        \caption{Example Responses}
        \label{tab:example}
        \begin{tabular}{ccp{0.7\textwidth}}
            \toprule
            Series & Direction & Headline/Response \\
            \midrule
            & &(1989-06-29) U.S. Reaches Accords Widening Access To the Mobile Phone Business in Japan \\
\midrule
S\&P &   1 &        The news signals potential growth opportunities for US mobile phone companies, which could positively impact the S\&P 500 index. \\
   UE &  -1 & Increased access to the Japanese mobile phone market will likely create new job opportunities in the U.S. telecommunications industry. \\
  CPI &  -1 &                      Increased competition and access to the Japanese mobile phone market may lead to lower prices for U.S. consumers. \\
            \\[10pt]
            \midrule
            & &(2001-02-28) Bush Offers Tax Cuts and Tight Budgets to Aid `Faltering' Economy \\
\midrule
S\&P &   1 &                           Tax cuts stimulate the economy, which could lead to increased corporate profits and higher stock prices. \\
   UE &   0 &                         Tax cuts may stimulate investment, but tight budgets could reduce government spending and slow job growth. \\
  CPI &   1 & Tax cuts usually stimulate spending, which can increase demand and raise prices, but budget tightening may counteract that effect. \\
            \\[-9pt]
            \bottomrule
        \end{tabular}
    \end{center}
{\footnotesize {\it Note.}
    Reports a sample of example article headlines and the corresponding responses from GPT.
    The first column reports the series queried.
    The second column reports the direction of the query.
    The third column reports the headline and the corresponding response from GPT.
    }
\end{table}

Finally, since GPT produces article level binary expectations, I aggregate these expectations to place them at a comparable frequency to the survey data.
To do this aggregation I compute a ``balance statistic'': the proportion of articles where GPT responds with increases minus the proportion articles where GPT responds with decreases:

\begin{equation}
    F^{gpt}_{t}(X^k_{t+h}) = \frac{\sum_{i \in A_{t}} \mathbb{I}(\text{Increase})^k_i - \mathbb{I}(\text{Decrease})^k_i}{\sum_{i \in A_{t}} \mathbb{I}(\text{Increase})^k_i + \sum_{i \in A_{t}} \mathbb{I}(\text{Decrease})^k_i}.
\end{equation}
This approach is common for a number of survey measures and is used for other surveys such as the Gallup survey, the University of Michigan Survey of Consumers, and the American Association of Individual Investors survey.

\FloatBarrier

\section{Return Expectations}

I first validate GPT's expectations of returns by comparing them to two publicly available benchmark return series used in \cite{Greenwood2014}.
The first is the American Association of Individual Investors (AAII) Investor Sentiment Survey.
The AAII survey is a weekly survey of members of the AAII running from 1987 up to the present day which measures the percentage of participants that are bullish, bearish, or neutral on the stock market for the next six months.
Following \cite{Greenwood2014}, I aggregate the weekly responses to monthly for the majority of my analysis.
The second survey is Duke or Graham and Harvey survey of chief financial officers (CFOs), started in 1998 by John Graham and Campbell Harvey.
The survey requests CFO views on a variety of macro and firm specific quantities including their expectations of returns for the U.S. stock market over the next twelve months.
The survey runs from 2000 to the present day.

Figure \ref{fig:retwsj} reports the time series of GPT's expectations using a three-month window for aggregation, along with the corresponding survey series.
Additionally, it reports the correlation between each survey series and various aggregates of GPT's expectations.
I consider aggregates over one, two, and three month windows as well as an exponentially weighted average over daily aggregates.
For the exponentially weighted average I report correlation for the optimal smoothing parameter $\lambda$ that maximizes the correlation with the survey series.

I find that for all different aggregation methods GPT's expectations are significantly correlated with the existing survey measures.
For the remainder of this section I focus on the three-month aggregation window.
For this window, the average correlation between GPT's expectations and the existing survey measures is 0.47.
For comparison \cite{Greenwood2014} find an average correlation of 0.43 among the full set of return expectation surveys they evaluate.

\begin{figure}[!t]
 \begin{center}
 \caption{Time Series of Return Expectations \label{fig:retwsj}}
 \vspace{0.1in}\footnotesize
\includegraphics[width=\textwidth]{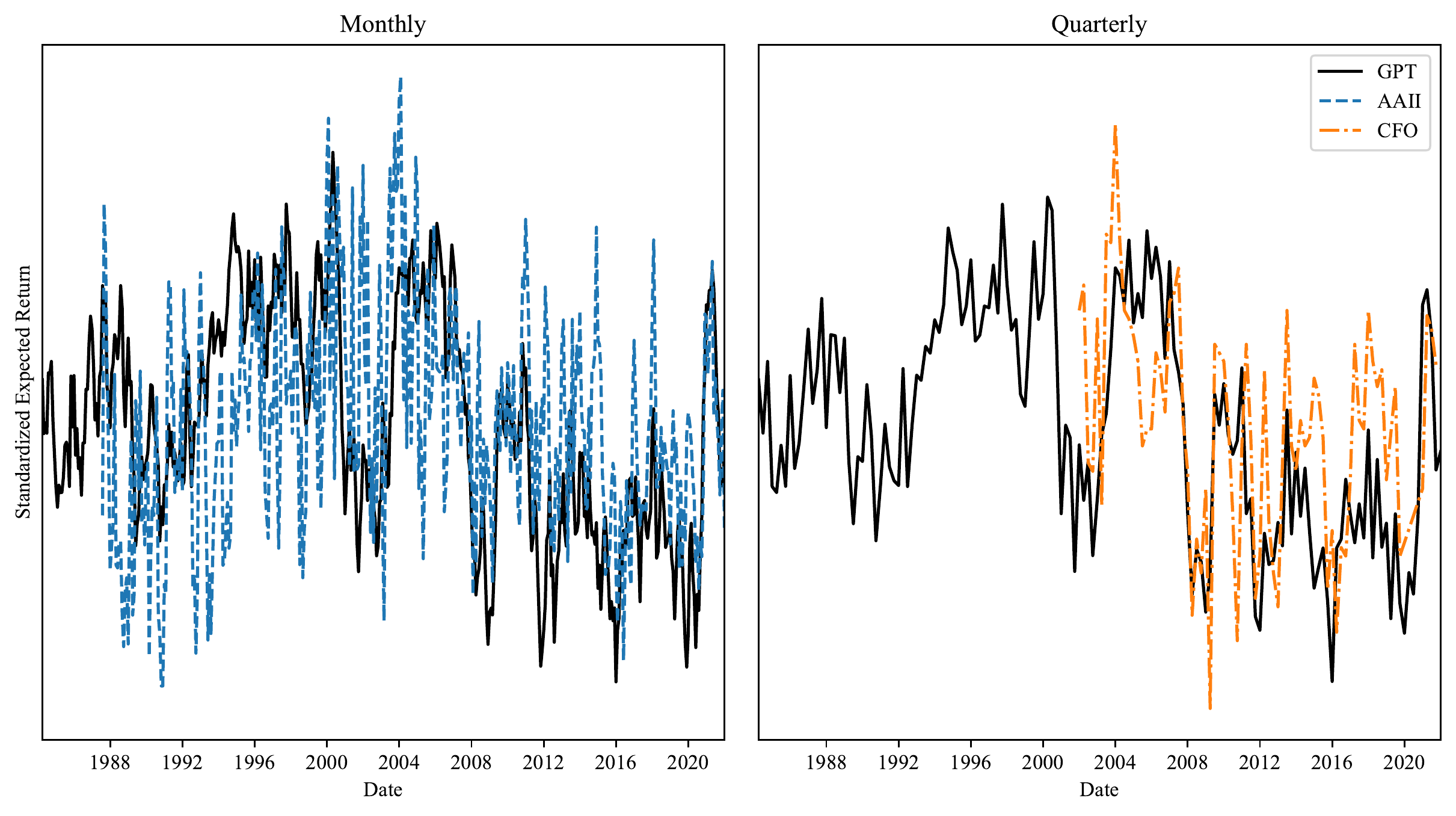}\\
    \label{tab:retcorr}
 \end{center}
    \begin{center}
    \begin{tabular}{lcccc}
    \toprule
        & one month avg. & two month avg. & three month avg. & opt. EWMA \\
    \midrule
        \input{ctab.tex}
        \\[-9pt]
        \bottomrule
    \end{tabular}
    \end{center}
{\footnotesize {\it Note.}
Reports the time-series of GPT's monthly/quarterly standardized expectations overlaid with the AAII and CFO surveys respectively.
Bottom table reports the correlation coefficients and corresponding $t$-stats in brackets.
$t$-stats use Newey-West standard errors with a 12 month lag.
one/two/three month avg. correspond to GPT expectations aggregated over one/two/three month windows respectively.
opt. EWMA corresponds to an exponentially weighted average with the optimal tuning parameter.
    }

\end{figure}

These results suggest GPT's expectations closely match the variation in existing survey measures, as a result I next evaluate how well GPT's expectations match key properties of existing return expectations.
To test this, I next compare the correlation between GPT's expectations and alternative measures studied previously in the literature vs. the existing survey measures.

An extensive literature in asset pricing has focused on the importance of extrapolative expectations (\cite{cutler1990speculative}, \cite{barsky1993does}, \cite{lakonishok1994contrarian}, \cite{Barberis2015Jan}, \cite{jin2022asset}).
As a result, I evaluate the correlation of different survey expectations against the past twelve-months returns ($R_{t-12}$) of the U.S. stock market, following \cite{Greenwood2014} and \cite{Nagel2022Feb}.
Additionally, a recent literature has leveraged the increased availability of realized trading behavior as a proxy for investor beliefs (\cite{Giglio2019Apr}, \cite{Gabaix2021Jun}, \cite{Alekseev2022Dec}).  Following this literature, I evaluate the correlation between different survey expectations and mutual fund flows into equities.  To compute this measure I combine the Thomson Reuters Mutual Fund Holdings S12 database with the CRSP Survivor-Bias-Free US Mutual Funds and follow the data cleaning procedure detailed in \cite{Alekseev2022Dec}.

An exhaustive literature in asset pricing has also studied empirical measures of objective expected returns.
Importantly, there exists a well known disconnect between these objective measures and subjective survey measures (\cite{Greenwood2014}, \cite{Nagel2022Feb}).
As a result, I next consider a number of objective expected return proxies.
First, I consider the log dividend-price ratio and 12 month changes in the log dividend-price ratio following much of the empirical literature.
Second, I consider the consumption wealth ratio (CAY) of \cite{Lettau2001Jun}.
Finally, I consider two expected return indices formed by regressing future 12-month returns on various sets of predictors.
In particular, I consider the same expected return index used in \cite{Greenwood2014} (following \cite{Fama1989Nov}) which uses the log dividend price ratio, the Treasury-bill yield, the default spread (the yield on BAA minus yield on AAA-rated bonds) and the term spread (the yield on ten-year government bonds minus the yield on three-month Treasury bill).
Additionally, I form an expected return index using the ``kitchen-sink'' predictors from \cite{Welch2008Jul} which represent a common benchmark in the empirical asset pricing literature.

\begin{figure}[htbp]
 \begin{center}
 \caption{Survey Correlations with Existing Moments \label{fig:retmom}}
\includegraphics[width=\textwidth]{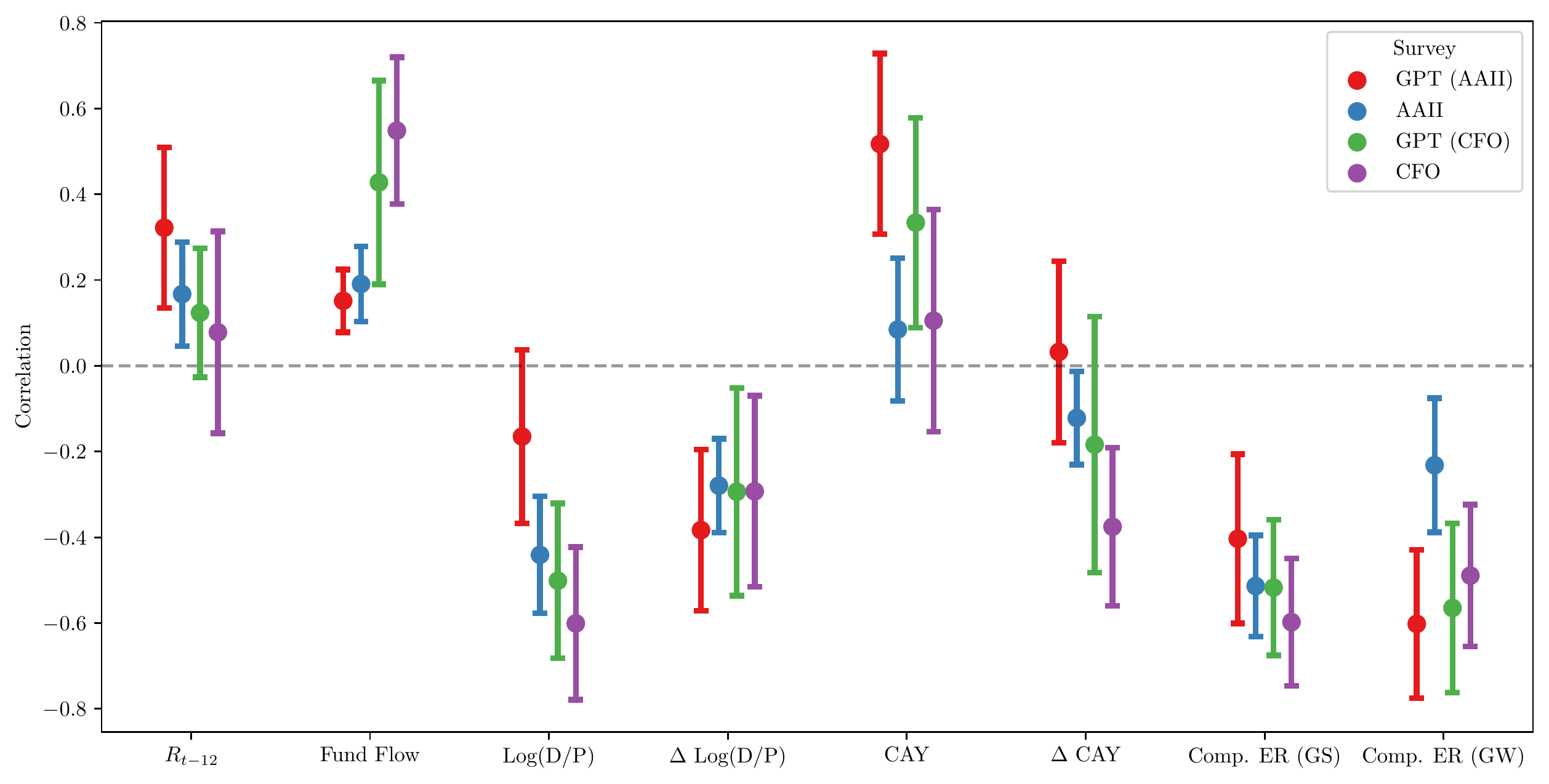}\\
 \end{center}
    \begin{center}
        \begin{tabular}{lcccccccc}
    \toprule
            & \multicolumn{2}{c}{GPT (AAII)} & \multicolumn{2}{c}{AAII} & \multicolumn{2}{c}{GPT (CFO)} & \multicolumn{2}{c}{CFO} \\
            & Corr. & $t$-Stat & Corr. & $t$-Stat & Corr. & $t$-Stat & Corr. & $t$-Stat \\
    \midrule
        \input{summary.tex}
        \\[-9pt]
        \bottomrule
    \end{tabular}
    \end{center}
{\footnotesize {\it Note.}
    Reports the correlation between each expectation series and the corresponding series and 90\% confidence intervals,
    standard errors are Newey-West with a 12 month lag.
    Bottom table reports the corresponding correlation coefficients and $t$-stats.
    $t$-stats use Newey-West standard errors with a 12 month lag.
    GPT (AAII) corresponds to the correlation with GPT's expectations at the monthly frequency for dates where the AAII survey is available.
    GPT (CFO) corresponds to the correlation with GPT's expectations at the quarterly frequency for dates where the CFO survey is available.
    $R_{t-12}$ corresponds to lagged 12 month returns.
    Fund Flow corresponds to the aggregate mutual fund flows into equities.
    Log(D/P) is the log dividend-price ratio and CAY the aggregate log consumption-wealth ratio of \cite{Lettau2001Jun}.
    $\Delta$ Log(D/P) and $\Delta$ CAY correspond to the respective 12-month changes.
    Comp. ER (GS) corresponds to the fitted values of a regression of one-year ahead returns on log dividend price ratio, the Treasury-bill yield, the default spread and the term spread ----- the composite expected return measure used in \cite{Greenwood2014}.
    Comp. ER (GW) corresponds to the fitted values of a regression of one-year ahead returns on the ``kitchen-sink'' predictors from \cite{Welch2008Jul}.
    }

\end{figure}

Figure \ref{fig:retmom} reports the correlations between each survey series and the corresponding measures discussed above.
For each survey series, I include a GPT based expectation series formed at the same frequency over the same time sample.
For all cases except changes in CAY and the GPT expectations proxy for AAII, GPT's correlations exhibit the same sign as the existing survey measures.
These results are evidence that GPT does not only match the variation in existing survey measures but more importantly the deviations from rational expectations noted in the literature, indicating the potential of these measures as a tool for studying nonrational expectations.

\begin{figure}[!t]
 \begin{center}
 \caption{Predictive Return Regressions\label{fig:cret}}
     \begin{tabular}{cc}
       Subjective ER & Objective ER \\
         \includegraphics[width=0.5\textwidth]{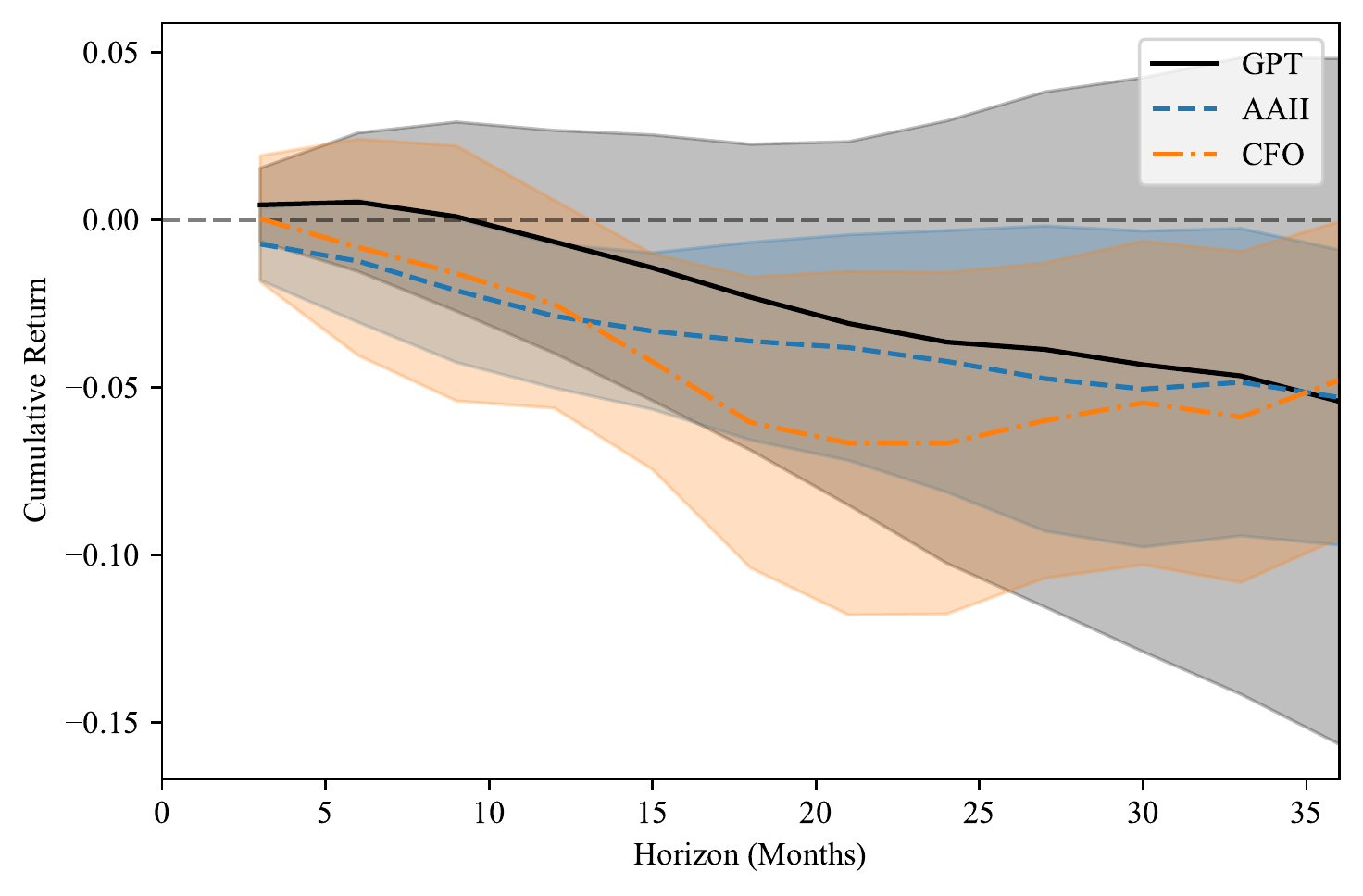} &
         \includegraphics[width=0.5\textwidth]{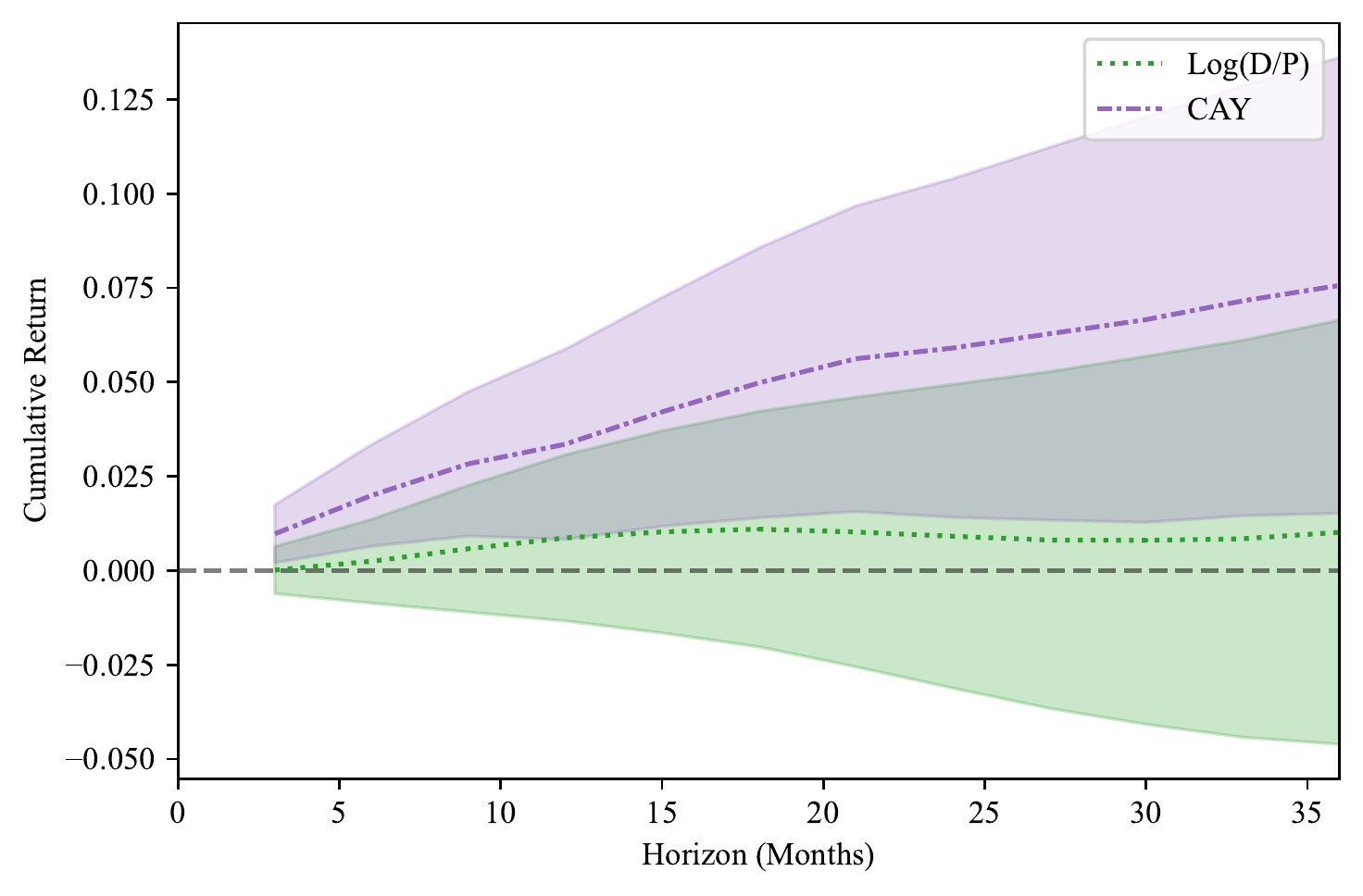} \\
     \end{tabular}
 \end{center}
{\footnotesize {\it Note.}
Reports the coefficients for a series of predictive regressions of future cumulative returns over the given horizon on subjective and objective expected return proxies.
Shaded bans report 90\% confidence intervals using Newey-West standard errors with the corresponding horizon as the number of lags.
    }
\end{figure}

Finally, it is the case that measures of expected returns should forecast future returns under models of rational expectations.
However, if anything, existing survey measures are negatively correlated with future returns.
As a result, I next compare the correlation of GPT's expectations with future returns to that of a number of alternative subjective and objective expected return measures.
Figure \ref{fig:cret} reports the correlation coefficients for a series of predictive regressions of future returns on the expectation measures discussed above.
GPT exhibits the same negative correlation with future returns as existing survey measures, which is in contrast to the positive correlation observed for objective measures.

\FloatBarrier

\section{Macroeconomic Expectations}
\label{sec:macro}

I next validate GPT's expectations by comparing them to a standard set of macroeconomic expectations: the Survey of Professional Forecasters (SPF).
The SPF is a quarterly survey of macroeconomic forecasts conducted by the Philadelphia Federal Reserve and remains the gold standard for work studying macroeconomic expectations (\cite{Coibion2012Feb}, \cite{Coibion2015Aug}, \cite{bordalo2020overreaction}, \cite{Angeletos2021Jan}, \cite{farmer2021learning}).
As professional forecasters, the respondents are some of the most informed agents in the economy and as such errors in their forecasts are notable.

The SPF covers a wide range of macroeconomic quantities during my sample, including CPI, real GDP, the unemployment rate and the federal funds rate.
I compare GPT's expectations to that of the SPF for the 13 series examined in \cite{Coibion2015Aug}.
For GPT's expectations I use a quarterly aggregate of the article level expectations over the quarter prior to the surveyed period.
The SPF releases forecasts at a variety of horizons, however, since the horizon of GPT's expectations is not known, I compare GPT's expectations to an average over the 1-4 quarter horizon forecasts.
My results are robust to this choice and results for each of the 1-4 quarter horizon forecasts are shown in appendix \ref{sec:horizon}.

\begin{figure}[hbtp]
 \begin{center}
 \caption{GPT/SPF Correlations \label{fig:spfcorr}}
 \vspace{0.1in}\footnotesize
\includegraphics[width=0.8\textwidth]{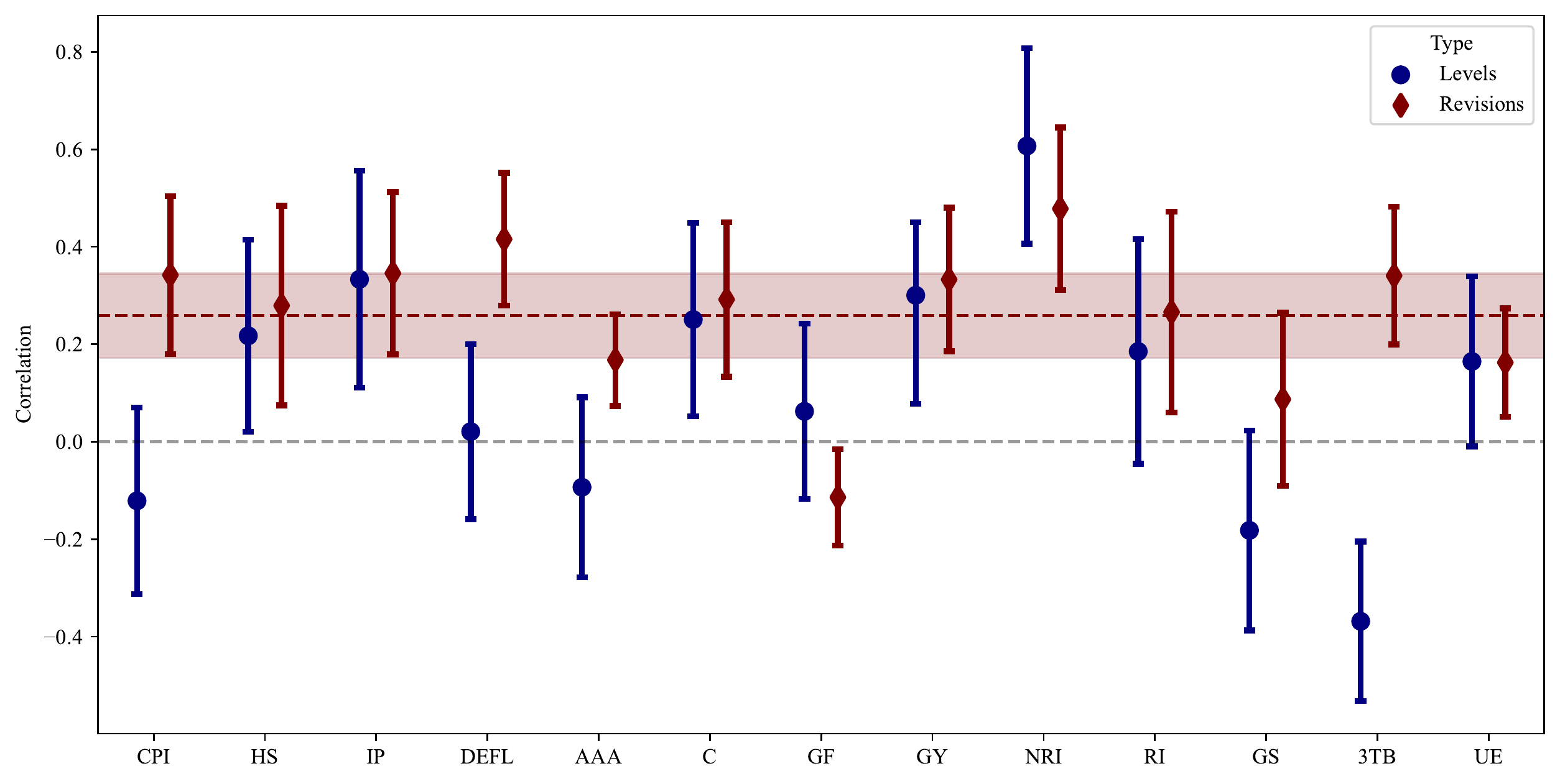}\\
 \end{center}
    \begin{center}
    \begin{tabular}{lcccc}
    \toprule
        & \multicolumn{2}{c}{Levels} & \multicolumn{2}{c}{Revisions} \\
        & Corr. & T-Stat. & Corr. & T-Stat. \\
    \midrule
        \input{lvl_rev.tex}
        \\[-9pt]
        \bottomrule
    \end{tabular}
    \end{center}
{\footnotesize {\it Note.}
Reports the correlation between levels and revisions of SPF expectations and GPT expectations.
Additionally reports 90\% confidences intervals for the correlation coefficients.
Dashed maroon line reports the panel correlation coefficient and shaded maroon band the corresponding 90\% confidence interval.
Standard errors for the single variable regressions are Newey-West, panel standard errors are Driscoll-Kraay.
Bottom table reports the corresponding correlation coefficients and $t$-stats.
    }
\end{figure}

I first evaluate how well GPT's expectations correlate with SPF forecasts.
I report results for both levels $F_t(X^k_{t+h})$ and revisions $F_t(X^k_{t+h}) - F_{t-1}(X^k_{t+h})$.
Revisions capture the new information embedded in the forecast and given GPT's expectations are formed from granular pieces of recent news it is likely that they will be more closely related to revisions than levels.
Figure \ref{fig:spfcorr} reports these correlations -- a full set of time-series plots of GPT's expectations vs. the corresponding SPF revisions is available in appendix \ref{sec:spfts}.
For all but two of the sampled series, federal government consumption and state and local government consumption, GPT's expectations significantly correlated with the corresponding revisions at the 90\% confidence level.

Much of the literature studying deviations from FIRE focuses on the predictability of forecast errors.
In particular, considerable focus is given to the rigidity of beliefs to new information measured using \cite{Coibion2015Aug} (CG) regressions.
CG regressions regress forecast errors on the corresponding revisions in expectations -- positive coefficients capture underreaction, while negative coefficients capture overreaction.

\begin{figure}[hbtp]
 \begin{center}
 \caption{Coibion-Gorodnichenko Regressions \label{fig:cgreg}}
 \vspace{0.1in}\footnotesize
\includegraphics[width=0.8\textwidth]{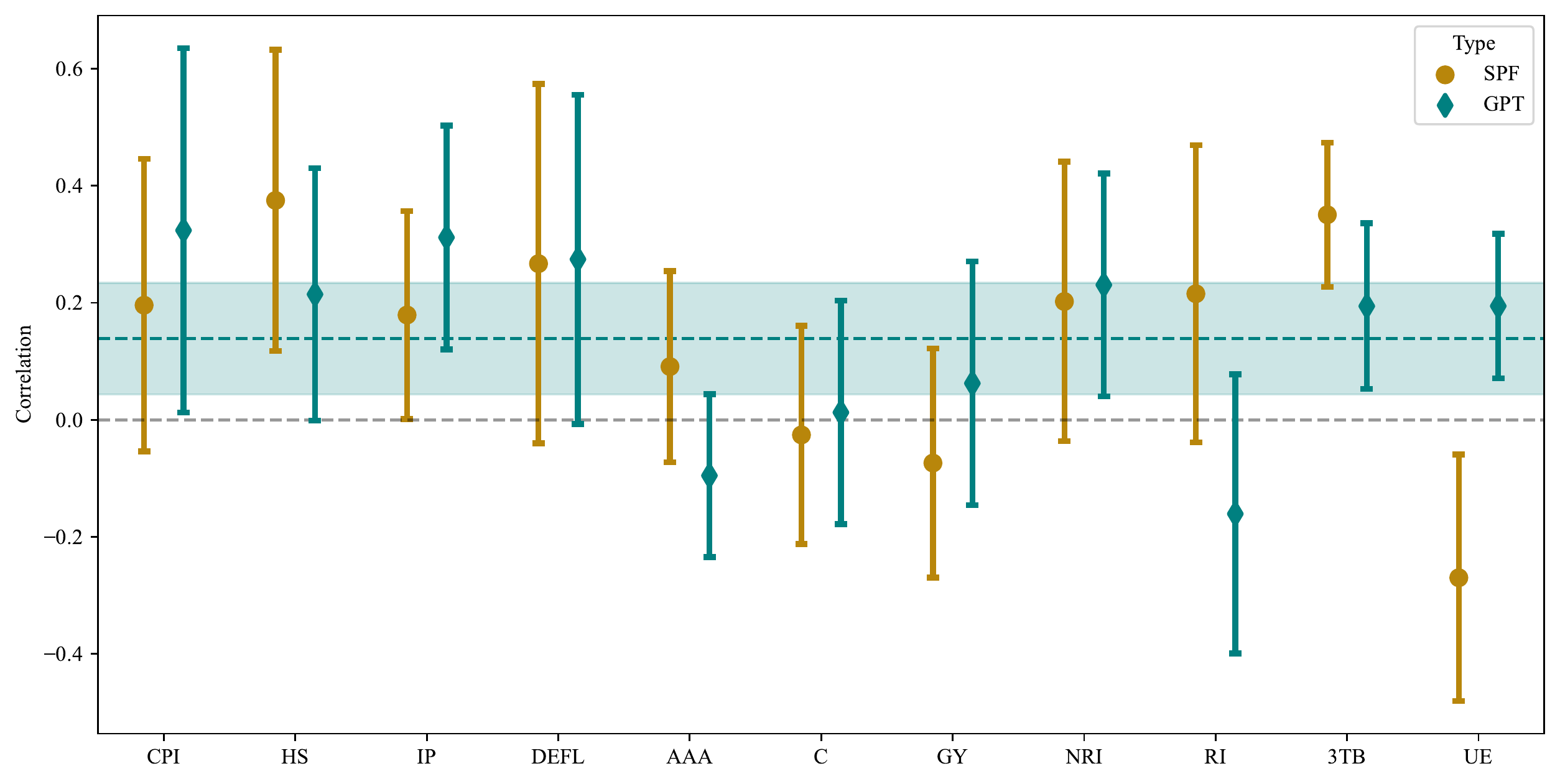}\\
 \end{center}
    \begin{center}
    \begin{tabular}{lcccc}
    \toprule
        & \multicolumn{2}{c}{SPF} & \multicolumn{2}{c}{GPT} \\
        & Corr. & T-Stat. & Corr. & T-Stat. \\
    \midrule
        \input{cg_inst.tex}
        \\[-9pt]
        \bottomrule
    \end{tabular}
    \end{center}
{\footnotesize {\it Note.}
Reports the CG coefficients for SPF revisions and GPT expectations.
Additionally reports 90\% confidences intervals for the coefficients.
Dashed teal line reports the panel CG coefficient and shaded teal band the corresponding 90\% confidence interval.
Standard errors for the single variable regressions are Newey-West, panel standard errors are Driscoll-Kraay.
Bottom table reports the corresponding correlation coefficients and $t$-stats.
    }
\end{figure}

Given GPT's significant correlation with SPF revisions, I next evaluate whether the well documented underreaction in macroeconomic expectations can be explained by the component of revisions associated with GPT.
To do this, for the series which exhibit a significant correlation between GPT and SPF revisions (that is, excluding government consumption), I regress SPF forecast errors on the corresponding version of GPT's expectations.
Figure \ref{fig:cgreg} reports the correlations for these regressions and compares them to the correlations using the original SPF revisions.
For all but two of the series, the correlation between GPT's expectations and SPF forecast errors is positive and the overall pooled results are positive and significant at the 90\% confidence level.\footnote{Unemployment exhibits significant overreaction in the original SPF revisions, going against the literature.  My sample includes the 2020 Covid-19 pandemic which may impact this result.  To test this, I re-estimate the same procedure limiting myself to the period prior to 2019.  These results are reported in appendix \ref{sec:covid}. The significant positive correlation between GPT's expectations and SPF forecast errors remains and the original CG results align with the literature.}

\FloatBarrier

\section{Memorization or Generalization?}

Given GPT's training period overlaps with my sample period I next evaluate whether GPT's expectations result from data leakage or memorization of the training sample.
The concern about memorization is different from standard concerns for overfitting.
LLMs are trained to predict the next token in a sequence and direct overfitting here would indicate prompts in the format I specify here exist in the training sample which seems unlikely.
However, GPT's training corpus could potentially contain information and discussions about the future values of the variables I'm interested in which it recites when prompted.
To address this concern I run two tests.
First, I extend my sample of \emph{WSJ} articles to include articles from after GPT's training period -- after September of 2021.
Second, I test whether GPT's expectations are driven by revisions in the SPF forecasts or by the realized values of the variables.

To evaluate GPT's out-of-sample expectations, I scrape all available \emph{WSJ} articles from the \emph{WSJ} archive from September 2021 to March 2023.
I then apply the cleaning procedure detailed in appendix \ref{app:filtersoos} and query GPT for its expectation for each article as with the main sample.
AAII return expectations are released weekly, and as a result this gives me 79 observations outside of GPT's training period.
For the SPF only five quarters of data are available -- as a result I run a panel regression of revisions in SPF expectations on GPT's expectations for the full set of series studied above.
Table \ref{tab:ooscurr} reports the results for these tests and indicates that the significant correlation I observe for my main sample is not the result of memorization.

\begin{table}[!h]
    \begin{center}
        \caption{Correlation with GPT OOS}
    \label{tab:ooscurr}
        \begin{tabular}{lcccccc}
    \toprule
            & AAII (Weekly) & SPF Avg. & SPF 1 & SPF 2 & SPF 3 & SPF 4 \\
    \midrule
        Correlation &    0.46 &    0.61 &    0.49 &    0.64 &    0.57 &    0.53 \\
T-Stat      &  [4.54] &  [6.11] &  [4.50] &  [6.61] &  [5.45] &  [4.93] \\
N           &      79 &      65 &      65 &      65 &      65 &      65 \\
        \\[-9pt]
        \bottomrule
    \end{tabular}
    \end{center}
{\footnotesize {\it Note.}
Reports the correlation between GPT's out-of-sample expectations and various benchmarks.
Each SPF column reports a different horizon and standard errors are Driscoll-Kraay.
    }
\end{table}

Next, I evaluate whether GPT's expectations are driven by revisions in the SPF forecasts or by the realized values of the variables.
To do this I run a panel regression of revisions in SPF expectations on GPT's expectations for the full set of series studied above.
Table \ref{tab:pcurr} reports the results for these tests and indicates that the significant correlation I observe for my main sample is not the result of GPT having knowledge about future realizations beyond the SPF expectations.

\begin{table}[!h]
    \begin{center}
        \caption{Correlation between GPT and SPF revisions and realized values}
    \label{tab:pcurr}
        \begin{tabular}{lccccc}
    \toprule
            Horizon & Avg. & 1 & 2 & 3 & 4\\
    \midrule
        SPF Revision &     0.25 &    0.20 &    0.17 &    0.16 &    0.14 \\
             &  [11.65] &  [8.92] &  [7.92] &  [7.02] &  [6.19] \\
Realized     &     0.02 &    0.01 &    0.02 &    0.02 &    0.01 \\
             &   [1.80] &  [0.47] &  [1.44] &  [1.36] &  [0.97] \\
$R^2$           &     6.71 &    4.07 &    3.27 &    2.53 &    1.95 \\
        \\[-9pt]
        \bottomrule
    \end{tabular}
    \end{center}
{\footnotesize {\it Note.}
Reports the correlation between GPT's expectations and SPF revisions and realized values.
Standard errors are Driscoll-Kraay.
    }
\end{table}

\FloatBarrier

\section{Conclusion}

In this paper I introduce a new methodology to form expectation proxies and study beliefs using large language models.
I find that the resulting expectation proxies exhibit many of the same deviations from full-information rational expectations as their current survey counterparts.
These results provide new evidence to help guide the development of models of expectations formation.

These facts impose new constraints on models of beliefs depending on the origins of these deviations.
If these deviations emerge from the estimated LLM itself then further work is needed to study the static behavior of these methods.
On the other hand, if the deviations emerge from the \emph{WSJ} articles themselves then further study of the narrative dynamics of these articles is needed.
Overall, there is still much to learn about LLMs as human simulacra but the results here indicate the potential of these methods to open a new agenda of study.


%
%

\clearpage

\bibliography{expectations}{}
\bibliographystyle{jf}

\clearpage

\appendix

\section{Constructing the \emph{WSJ} Samples}

\subsection{Constructing the \emph{WSJ} 1984-2021 Sample}
\label{app:filtersmain}

I conduct data processing steps in the following order:
\begin{enumerate}
\item Remove all articles prior to January 1984 and after December 2021.
\item Exclude articles with page-citation tags corresponding to any sections other than A, B, C, or missing.
\item Exclude articles corresponding to weekends.
\item Exclude articles with subject tags associated with obviously non-economic content such as sports.  List of exclusions available from authors on request.
\item Exclude articles with the certain headline patterns (such as those associated with data tables or those corresponding to regular sports, leisure, or books columns).  List of exclusions available from authors on request. 
\item Exclude articles with less than 100 words.
\item Exclude articles with headlines less than 10 words.
\item Sample 300 articles for each month of data.
\end{enumerate}

\subsection{Constructing the \emph{WSJ} Oct 2021-Mar 2023 Sample}
\label{app:filtersoos}

I conduct the following data processing steps in the following order:
\begin{enumerate}
    \item Exclude articles with headlines less than 8 words.
    \item Exclude articles not in the following sections: ``U.S'', ``BUSINESS'', ``WORLD'', ``POLITICS'', ``WSJ NEWS EXCLUSIVE'', ``HEARD ON THE STREET'', ``FINANCE'', ``EARNINGS'', ``MARKETS'', ``U.S. MARKETS'', ``U.S. ECONOMY'', ``ECONOMY''.
\end{enumerate}

\FloatBarrier

\section{Coocurrence of GPT Responses}
\label{app:coocurr}

This section reports the coocurrence between the article level increase/decrease responses for the GPT expectations series studied.

\begin{figure}[!t]
 \begin{center}
 \caption{Coocurrence Matrix \label{fig:coocurr}}
 \vspace{0.1in}\footnotesize
\includegraphics[width=\textwidth]{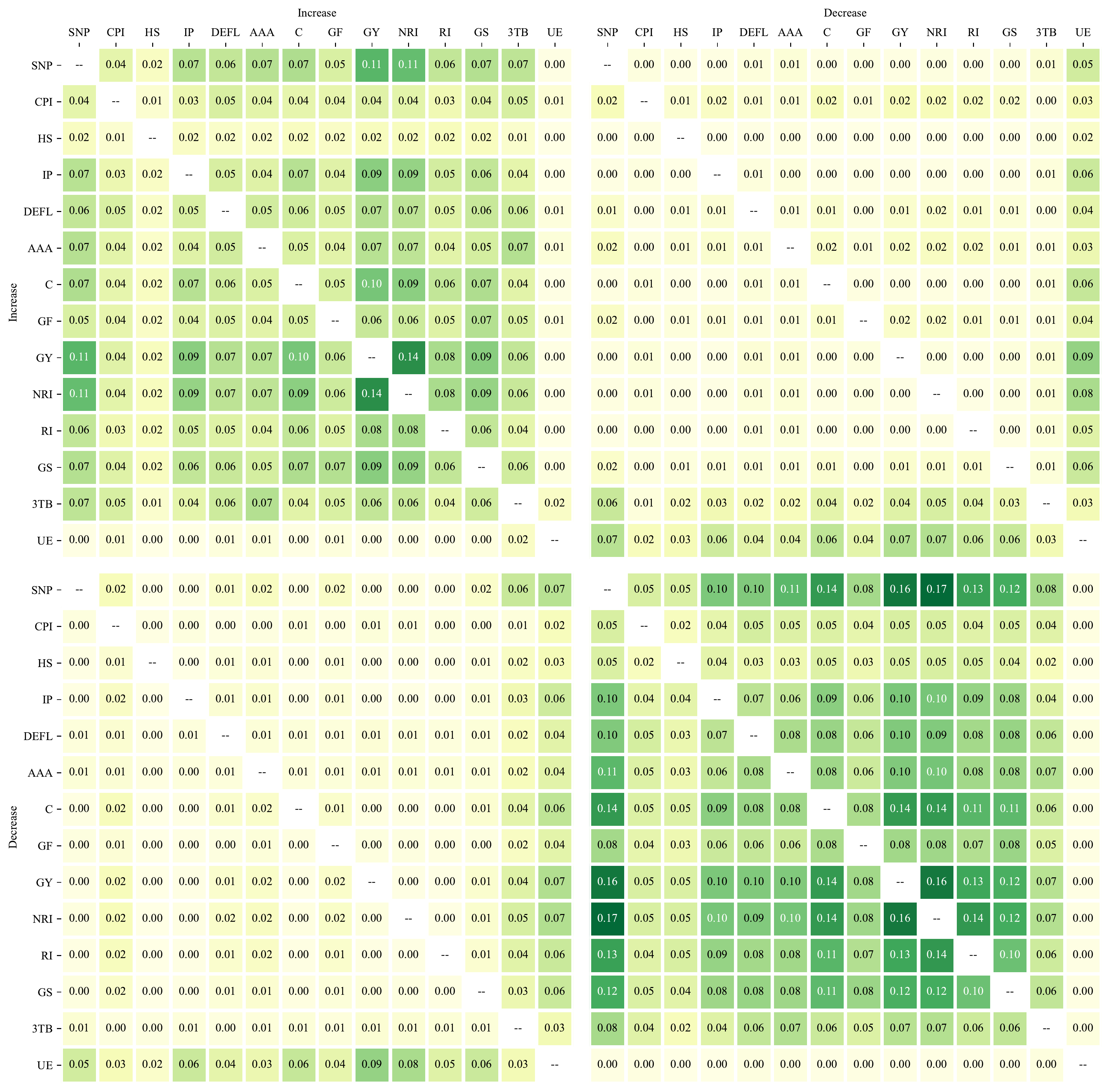}\\
 \end{center}
{\footnotesize {\it Note.}
    Reports the cooccurence proportion of each indicator for different surveyed series.
}
\end{figure}

\FloatBarrier

\section{Time-Series of GPT/SPF Expectations}
\label{sec:spfts}

This section reports time series plots the GPT expectations and average revisions for the SPF series studied in section \ref{sec:macro}.

\begin{figure}[!h]
 \begin{center}
 \caption{Time Series of GPT Expectations and SPF Revisions \label{fig:retwsj}}
 \vspace{0.1in}\footnotesize
\includegraphics[width=\textwidth]{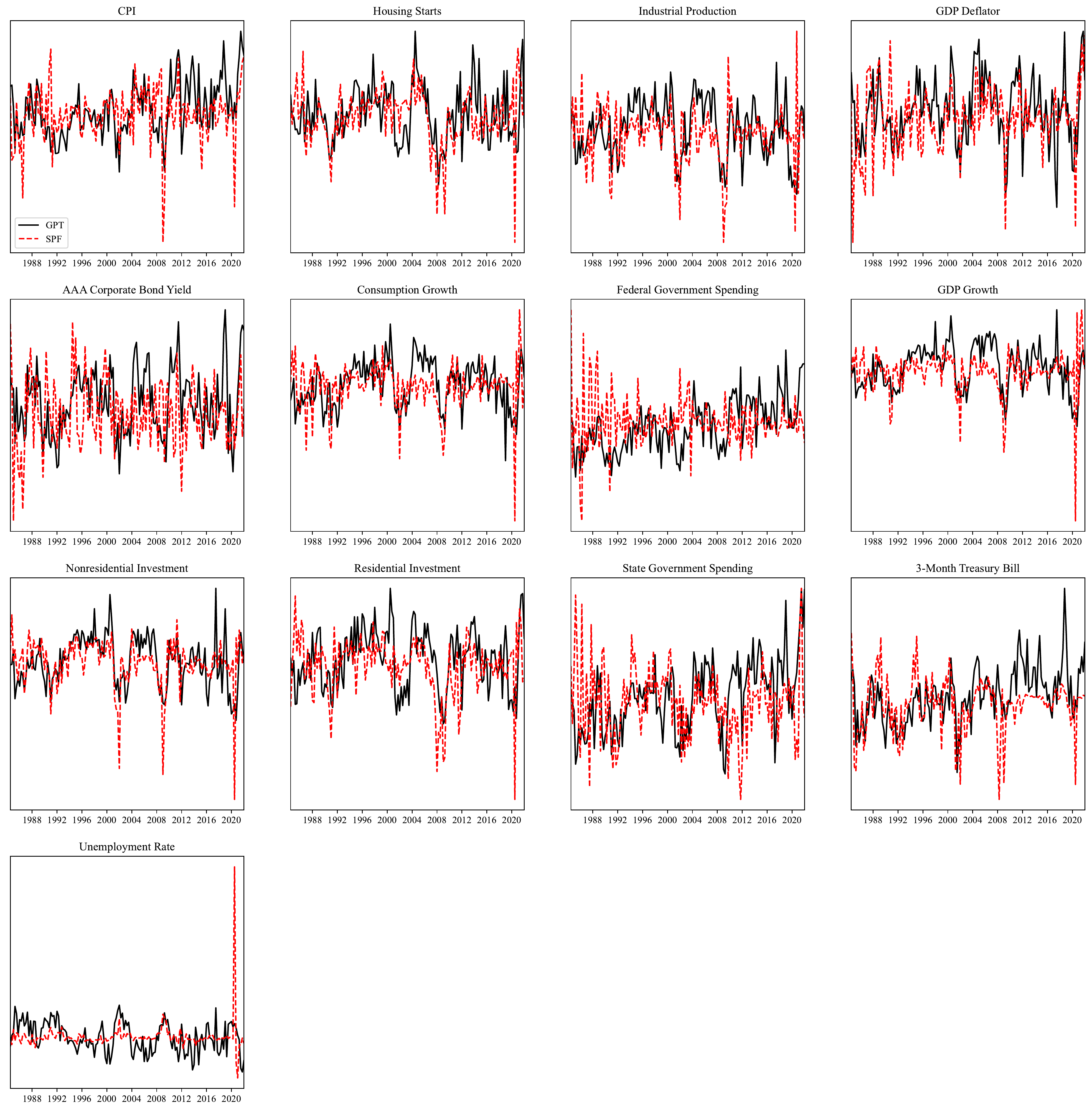}\\
 \end{center}
{\footnotesize {\it Note.}
Reports the time series of GPT expectations (black) and corresponding SPF revisions (red).
    }
\end{figure}

\FloatBarrier

\section{GPT/SPF Correlations Over Different Horizons}
\label{sec:horizon}

This section reports results comparable to section \ref{sec:macro} but partitioned over different horizons.

\begin{figure}[!h]
 \begin{center}
 \caption{SPF Revision Correlation Across Horizons \label{fig:spfhorizon}}
 \vspace{0.1in}\footnotesize
\includegraphics[width=\textwidth]{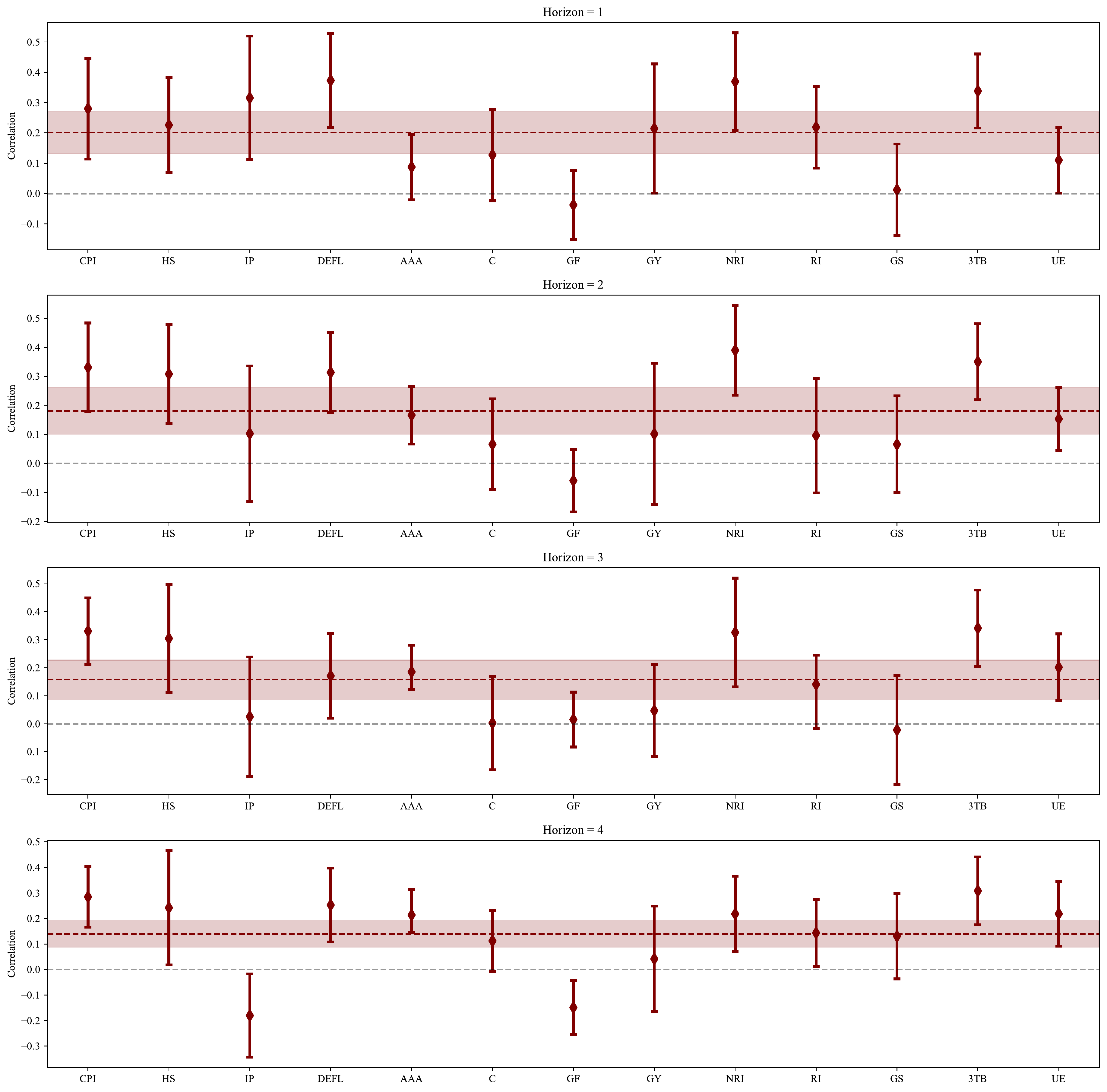}\\
 \end{center}
{\footnotesize {\it Note.}
Reports the correlation between revisions of SPF expectations and GPT expectations for different horizons.
Additionally reports 90\% confidences intervals for the correlation coefficients.
Dashed maroon line reports the panel correlation coefficient and shaded maroon band the corresponding 90\% confidence interval.
Standard errors for the single variable regressions are Newey-West, panel standard errors are Driscoll-Kraay.
    }
\end{figure}

\begin{figure}[!h]
 \begin{center}
 \caption{Coibion-Gorodnichenko Coefficients Across Horizons \label{fig:cghorizon}}
 \vspace{0.1in}\footnotesize
\includegraphics[width=\textwidth]{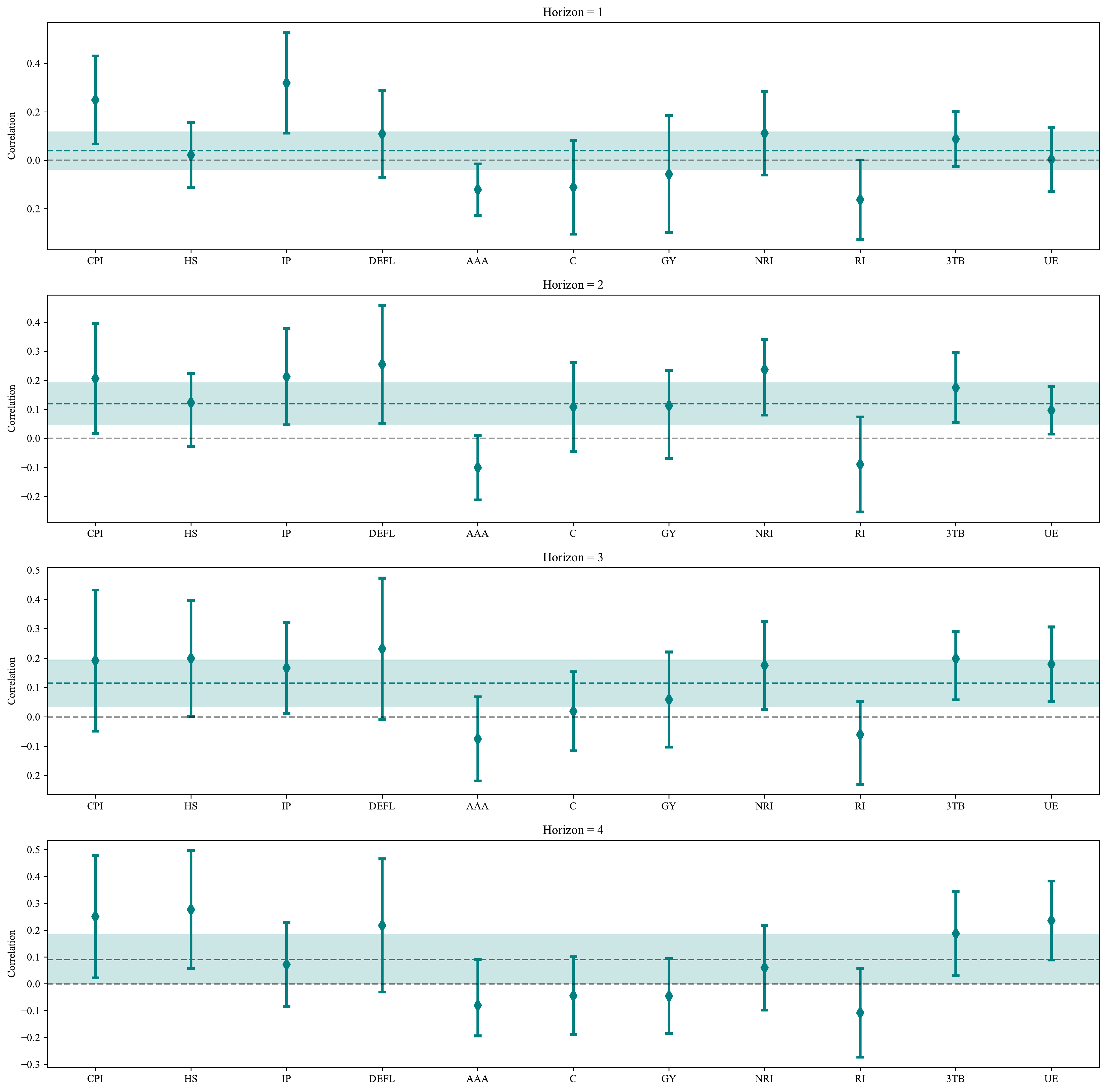}\\
 \end{center}
{\footnotesize {\it Note.}
Reports the CG coefficients for SPF revisions and GPT expectations across different horizons.
Additionally reports 90\% confidences intervals for the coefficients.
Dashed teal line reports the panel CG coefficient and shaded teal band the corresponding 90\% confidence interval.
Standard errors for the single variable regressions are Newey-West, panel standard errors are Driscoll-Kraay.
    }
\end{figure}

\FloatBarrier

\section{Covid}
\label{sec:covid}

This section reports results comparable to \ref{sec:macro} but excludes all data after 2019 to avoid the effects of the Covid-19 pandemic.

\begin{figure}[!h]
 \begin{center}
 \caption{GPT/SPF Correlations (Pre 2019)\label{fig:spfcovid}}
 \vspace{0.1in}\footnotesize
\includegraphics[width=0.8\textwidth]{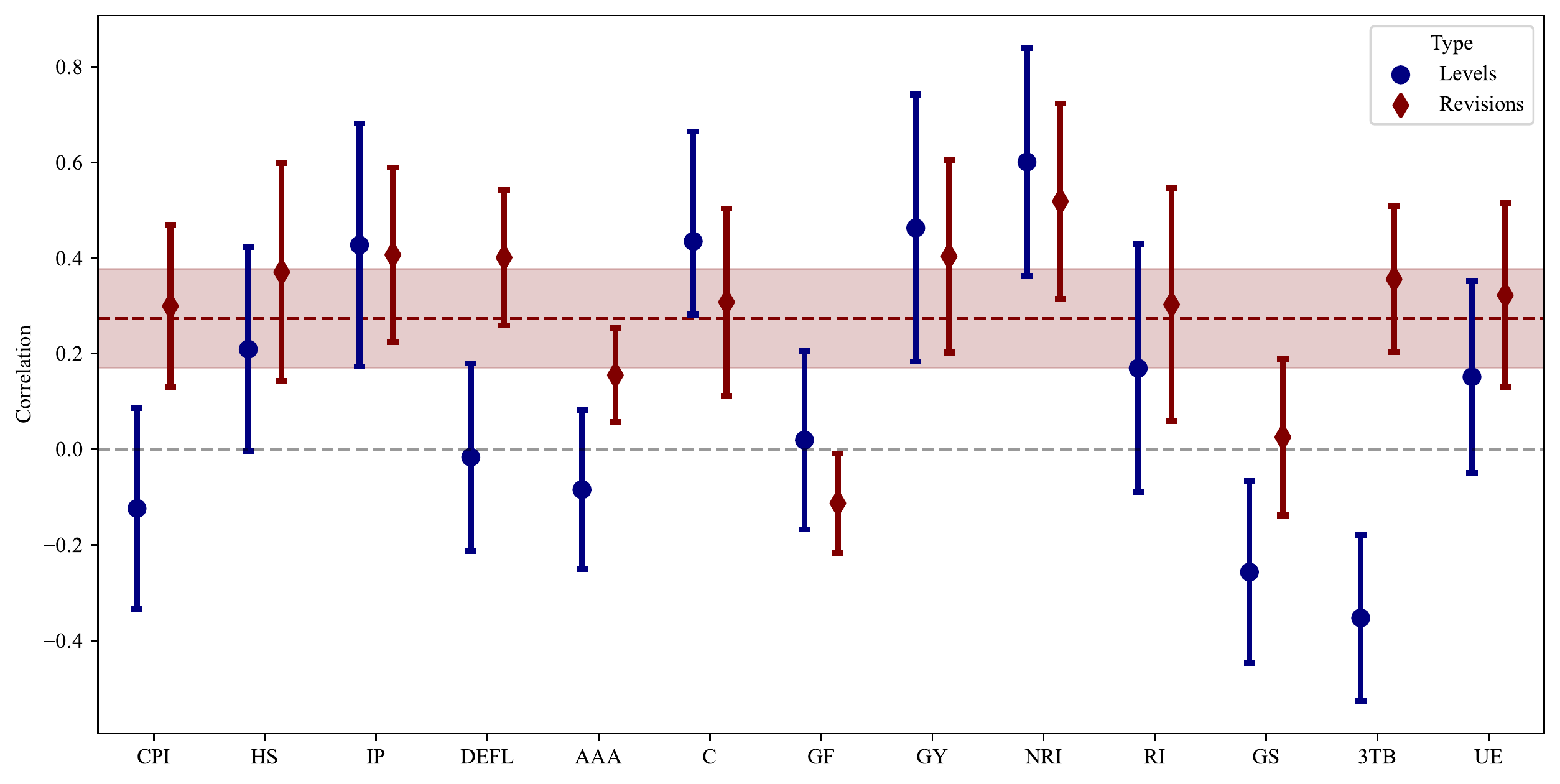}\\
 \end{center}
    \begin{center}
    \begin{tabular}{lcccc}
    \toprule
        & \multicolumn{2}{c}{Levels} & \multicolumn{2}{c}{Revisions} \\
        & Corr. & T-Stat. & Corr. & T-Stat. \\
    \midrule
        \input{lvl_rev_covid.tex}
        \\[-9pt]
        \bottomrule
    \end{tabular}
    \end{center}
{\footnotesize {\it Note.}
Reports the correlation between levels and revisions of SPF expectations and GPT expectations.
Additionally reports 90\% confidences intervals for the correlation coefficients.
Dashed maroon line reports the panel correlation coefficient and shaded maroon band the corresponding 90\% confidence interval.
Standard errors for the single variable regressions are Newey-West, panel standard errors are Driscoll-Kraay.
Bottom table reports the corresponding correlation coefficients and $t$-stats.
    }
\end{figure}

\begin{figure}[!h]
 \begin{center}
 \caption{Coibion-Gorodnichenko Regressions (Pre 2019) \label{fig:cgcovid}}
 \vspace{0.1in}\footnotesize
\includegraphics[width=0.8\textwidth]{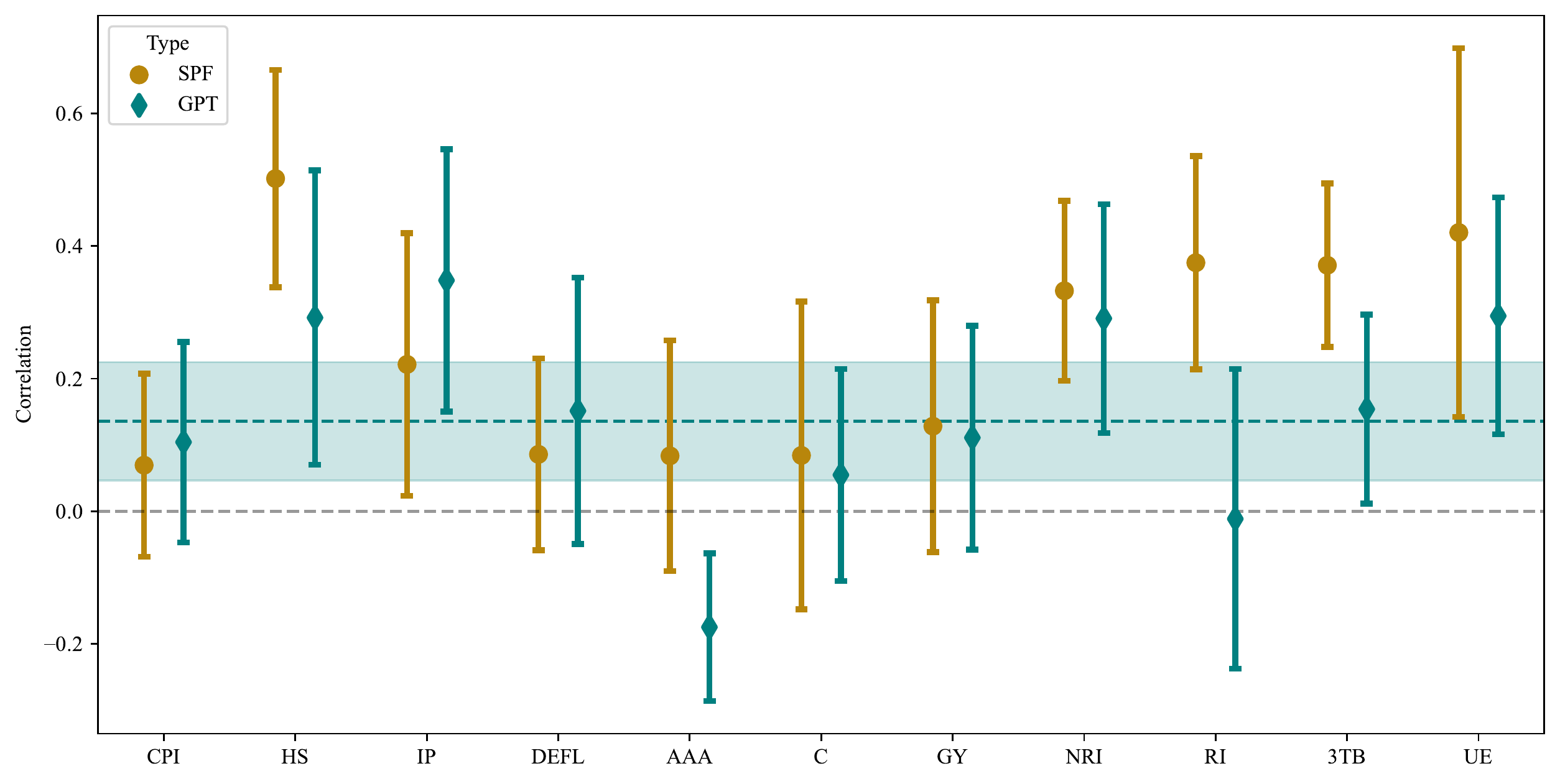}\\
 \end{center}
    \begin{center}
    \begin{tabular}{lcccc}
    \toprule
        & \multicolumn{2}{c}{SPF} & \multicolumn{2}{c}{GPT} \\
        & Corr. & T-Stat. & Corr. & T-Stat. \\
    \midrule
        \input{cg_inst_covid.tex}
        \\[-9pt]
        \bottomrule
    \end{tabular}
    \end{center}
{\footnotesize {\it Note.}
Reports the CG coefficients for SPF revisions and GPT expectations.
Additionally reports 90\% confidences intervals for the coefficients.
Dashed teal line reports the panel CG coefficient and shaded teal band the corresponding 90\% confidence interval.
Standard errors for the single variable regressions are Newey-West, panel standard errors are Driscoll-Kraay.
Bottom table reports the corresponding correlation coefficients and $t$-stats.
    }
\end{figure}

\end{document}